\documentclass[iop,apjl]{emulateapj}
\usepackage[utf8]{inputenc}
\usepackage{graphicx}
\usepackage{ifthen}
\usepackage{hyperref}
\newcommand {\apgt} {\ {\raise-.5ex\hbox{$\buildrel>\over\sim$}}\ }
\newcommand {\aplt} {\ {\raise-.5ex\hbox{$\buildrel<\over\sim$}}\ } 

\def\glc{{\sc Galacticus}}

\newcounter{IGMDone}
\def\IGM{\ifthenelse{\equal{\arabic{IGMDone}}{0}}{intergalactic medium (IGM)\setcounter{IGMDone}{1}}{IGM}}

\newcounter{ISMDone}
\def\ISM{\ifthenelse{\equal{\arabic{ISMDone}}{0}}{interstellar medium (ISM)\setcounter{ISMDone}{1}}{ISM}}

\newcounter{ICMDone}
\def\ICM{\ifthenelse{\equal{\arabic{ICMDone}}{0}}{intracluster medium (ICM)\setcounter{ICMDone}{1}}{ICM}}

\begin{document}

\title{Trends in Dwarf Early-Type Kinematics with Cluster-centric Radius Driven By Tidal Stirring}

\author{A. J. Benson\altaffilmark{1}}
\email{abenson@obs.carnegiescience.edu}

\author{E. Toloba\altaffilmark{1,2}}

\author{L. Mayer\altaffilmark{3,4}}

\author{J. D. Simon\altaffilmark{1}}

\author{P. Guhathakurta\altaffilmark{2}}
\affil{$^1$Observatories of the Carnegie Institution for Science, 813 Santa Barbara Street, Pasadena, CA 91101, USA}
\affil{$^2$UCO/Lick Observatory, University of California, Santa Cruz, 1156 High Street, Santa Cruz, CA 95064, USA}
\affil{$^3$Institute for Astronomy, ETH Z\"urich, Wolgang-Pauli-Strasse 27, 8093 Zurich, Switzerland}
\affil{$^4$Institute for Theoretical Physics, University of Z\"urich, Winterthurerstrasse 190, CH-9057 Zurich, Switzerland}

\begin{abstract}

We model the dynamics of dwarf early-type galaxies in the Virgo cluster when subject to a variety of environmental processes. We focus on how these processes imprint trends in dynamical state (rotational vs. pressure support as measured by the $\lambda^*_{\rm Re/2}$ statistic) with projected distance from the cluster center, and compare these results to observational estimates. We find a large scatter in the gradient of $\lambda^*_{\rm Re/2}$ with projected radius. A statistical analysis shows that models with no environmental effects produce gradients as steep as those observed in none of the 100 cluster realizations we consider, while in a model incorporating tidal stirring by the cluster potential 34\% of realizations produce gradients as steep as that observed. Our results suggest that tidal stirring may be the cause of the observed radial dependence of dwarf early-type dynamics in galaxy clusters.
\end{abstract}


\section{Introduction}

The strong morphological segregation found in galaxies located in regions of different density suggests that the environment plays a key role in their formation and evolution \citep[e.g][]{Dress80}. However, the physics involved in those processes, and which ones dominate over others, are questions that are not fully understood. Dwarf galaxies, systems with low-luminosities ($M_B \ge -18$) and shallow potential wells, are the ideal laboratories to test the role of environmental physics.

Within the class of dwarf galaxies, dwarf early-types (dEs) outnumber any other galaxy type in clusters \citep{Sandage85,Binggeli88}. Smooth and simple in appearance, they have been proven to be a very heterogeneous population showing different signatures that could be remnants of environmental mechanisms.
Detailed photometric studies of dEs in the Fornax, Virgo, and Coma clusters found spiral and irregular substructures in many of them \citep[e.g.][]{Jerjen00,Barazza02,Geha03,Graham03,deRijcke03,Lisker06,Ferrarese06,Janz12,janz_near-infrared_2013}. These features seem to indicate that dEs could be spiral and irregular galaxies faded by environmental effects. But, if dEs are old late-type galaxies, then, their dynamics should look alike, and dEs should be rotating systems. Due to its proximity and richness in number of objects, the Virgo cluster has been the main target for spectroscopic studies of dwarf early-type galaxies. The stellar kinematics of Virgo dEs show that the rotation speed changes from dE to dE even though those analyzed so far have similar brightnesses \citep[$-18 \le M_B \aplt -16$;][]{Pedraz02,Simien02,Geha02,Geha03,vanZee04,Chilingarian09,Toloba09,Toloba11,Rys13}, which suggests that the origin of this population must be more complicated, involving processes that, apart from aging the stellar population, change their dynamics.

\citet{Toloba09} found the first hint of a correlation between the strength of the rotation ($\lambda^*_{\rm R}$; a measure of the relative contribution of rotational and pressure support, including a correction for the ellipticity of the galaxy, as defined by \citealt{cappellari_sauron_2007}) and the distance to the center of the cluster. This result has been recently confirmed \citep{toloba_stellar_2014a} using the largest spectroscopic survey of spatially resolved kinematics for Virgo dEs observed to date \citep{toloba_stellar_2014}. This trend for pressure supported dEs (those whose dynamics are dominated by dispersion, low $\lambda^*_{\rm R}$) preferentially being concentrated in the center of the Virgo cluster, and the rotationally supported ones (dynamically dominated by rotation, high $\lambda^*_{\rm R}$) being found mainly in the outer parts of the cluster, puts strong constraints on the mechanisms shaping these galaxies. 

The environmental processes responsible for the transformation of late-type galaxies into dEs can be of two classes: gravitational tidal heating \citep[such as galaxy harassment;][]{Moore98,mastropietro_morphological_2005}, and hydrodynamical interaction with the intracluster medium \citep[such as ram pressure stripping;][]{LinFaber83,Boselli08}. These mechanisms predict different dynamics in the resulting systems. While galaxy harassment is fairly violent in that it heats the system and significantly reduces rotation, ram pressure stripping only removes the gas and leaves the dynamics of the stars untouched. The simulations that describe these scenarios are performed on a galaxy-by-galaxy basis, which makes it difficult to understand the radial trends in the galaxy dynamics found in clusters.

To go beyond single galaxy simulations, in this work we utilize a semi-analytic model of galaxy formation to investigate which environmental process (or processes) is responsible for the radial trends of $\lambda^*_{\rm R}$ observed in the Virgo cluster. We describe our modelling of this process in Section~\ref{sec:Model}, present out results in Section~\ref{sec:Results}, and draw our conclusions in Section~\ref{sec:Conclusions}

\section{Modelling}\label{sec:Model}

We use the semi-analytic galaxy formation code, \glc\footnote{\href{http://sites.google.com/site/galacticusmodel/}{\tt http://sites.google.com/site/galacticusmodel}; specifically version 0.9.3, revision 2046.}, to model the dynamical evolution of satellite galaxies in a Virgo cluster analog. While our semi-analytic model does not permit full dynamical modelling of galaxies, here we are interested only in the ratio of rotational support to pressure support. We therefore make use of the fact that \glc\ resolves each galaxy into a rotationally supported disk component and a dispersion supported spheroid component. Unlike the traditional use of disk and spheroid to define morphologically distinct components in semi-analytic models, here we use them to define dynamically distinct components. For clarity, we will therefore refer to ``rotation-supported'' and ``dispersion-supported'' components from here on.

To estimate the dynamical state of galaxies we use the $\lambda^*_{\rm R}$ estimator \citep{cappellari_sauron_2007}: 
\begin{equation}
\lambda^*_{\rm R} = {\int_0^R 2 \pi r^\prime \Sigma(r^\prime) V(r^\prime) {\rm d}r^\prime \over \int_0^R 2 \pi r^\prime \Sigma(r^\prime) \sqrt{\sigma^2(r^\prime)+V^2(r^\prime)} {\rm d}r^\prime},
\label{eq:lambdaR}
\end{equation}
where $\Sigma(r)$ is the projected surface density (in H-band light) of the galaxy at radius $r$, $\sigma^2(r)$ is the measured velocity dispersion and $V(r)$ the measured rotation speed. Assuming that the spheroid component is purely dispersion dominated with velocity dispersion $\sigma_{\rm s}(r)$, and that rotation is present in only the disk component with rotation curve $V_{\rm d}(r)$ then we can model the velocity distribution, $P(V)$, at $r$ as the sum of a Gaussian of width $\sigma_{\rm s}(r)$ and normalized area $\Sigma_{\rm s}(r)$, and a delta function at $V_{\rm d}(r)$ with normalized area $\Sigma_{\rm d}(r)$. The measured rotation speed is then:
\begin{equation}
V(r) = {\int_{-\infty}^{+\infty} P(V) V {\rm d}V \over \int_{-\infty}^{+\infty} P(V) {\rm d}V }= {\Sigma_{\rm d}(r) V_{\rm d}(r) \over [\Sigma_{\rm d}(r)+\Sigma_{\rm s}(r)]},
\end{equation}
and the measured velocity dispersion is:
\begin{eqnarray}
\sigma^2(r) &=& { \int_{-\infty}^{+\infty} P(V) [V-V(r)]^2 {\rm d}V \over \int_{-\infty}^{+\infty} P(V) {\rm d}V} \nonumber \\
&=& {  \Sigma_{\rm s}(r) [\sigma_{\rm s}^2(r)] + \Sigma_{\rm d}(r) [V_{\rm d}(r)-V(r)]^2  \over [\Sigma_{\rm d}(r)+\Sigma_{\rm s}(r)]}.
\end{eqnarray}
To match the observational analysis, the upper limits for the integrals in eqn.~(\ref{eq:lambdaR}) are taken to be half of the H-band half-light radius\footnote{While we choose to measure the degree of rotational support within Re$/2$ to match what was done in the observational analysis, in our models for physical morphological transformation mechanisms described below we will make use of the 3-D half-mass radii of galaxies in computing the strengths of various physical mechanisms, as these radii are better indicators of the overall structure of each galaxy.}---we label this quantity $\lambda^*_{\rm Re/2}$. Note that $\lambda^*_{\rm Re/2}$ when measured from observational data includes a correction for ellipticity. When measured from our model this is unecessary as we can choose to ``observe'' all model galaxies edge-on.

We explore several different models, differing in the environmental physics incorporated, to explore the role of each mechanism in driving radial trends in $\lambda^*_{\rm Re/2}$. For each model, we generate 100 realizations of a Virgo cluster analog\footnote{The cluster virial mass is distributed in the range $4.0$--$4.4\times10^{14}M_\odot$ \citep{mclaughlin_evidence_1999}, the merger history is generated using the algorithm of \protect\citeauthor{parkinson_generating_2008}~(\protect\citeyear{parkinson_generating_2008}; resolving progenitor halos down to $5\times10^9M_\odot$). Halo profiles are assumed to follow the form given by \protect\cite{navarro_universal_1997}, with concentrations set using the algorithm of \protect\cite{gao_redshift_2008}.} at $z=0$. The full merging history of each realization is constructed and the physics of galaxy formation solved within that merging hierarchy following the general methodology described by \cite{benson_galacticus:_2012}\footnote{The specific parameter set used to generate the models is available at \protect\href{http://users.obs.carnegiescience.edu/abenson/galacticus/parameters/dwarfEllipticals.tar.bz2}{\tt http://users.obs.carnegiescience.edu/abenson/\\galacticus/parameters/dwarfEllipticals.tar.bz2}}.

Of particular importance for this work are environmental effects acting upon satellite galaxies. We describe below simple models for all effects considered. Where necessary, the strength of such effects are evaluated at the pericenter of a satellite's orbit within its host halo (where the strength is maximized). Satellite orbits are chosen at random at infall from the cosmological distribution of \cite{benson_orbital_2005}. Given the energy and angular momentum of the satellite, and the gravitational potential of the host halo, the pericentric distance can be computed at any time. Since the host halo grows with time, the satellite orbit will change with time. We assume that the satellite orbital angular momentum is unchanged by growth of the host halo (which will be approximately true if the host halo is spherical and the potential grows slowly; \citealt{blumenthal_formation_1984}), while the energy of the satellite simply changes at a rate to maintain the same apocenter\footnote{We define the gravitational potential of our halos to be zero at infinity.}. We ignore the effects of dynamical friction, which will be weak for the satellites of interest due to their low mass relative to that of their host halo \citep{chandrasekhar_dynamical_1943,lacey_merger_1993}.

We consider the following five models:
\begin{description}
 \item [Minimal environmental effects:] This model contains the minimal possible complement of environmental effects. Specifically, the only difference between a satellite and central galaxy is that a satellite's dark matter halo is no longer growing so is no longer able to accrete new material from the \IGM. All remaining models add a single environmental effect to this baseline model. The dispersion-dominated component in this model can arise only from mergers occuring prior to a galaxy becoming a satellite;
 \item [Strangulation:] The hot atmosphere surrounding satellites is stripped away by ram pressure from the \ICM\ of their host halo at a rate computed using the model of \cite{font_colours_2008}. This will reduce the amount of new star formation in the rotation-supported component of satellites and so may affect $\lambda^*_{\rm Re/2}$;
 \item [Ram pressure stripping:] The \ISM\ of the galaxy is stripped away via ram pressure. The ram pressure force, $F_{\rm rp}$, is computed using the model of \cite{font_colours_2008}. We adopt a simple model to compute the mass loss rate due to ram pressure:
\begin{equation}
 \dot{M}_{\rm ISM} = - M_{\rm ISM}/\tau_{\rm rp},
\end{equation}
where
\begin{equation}
 \tau_{\rm rp} = \tau_{\rm dyn} [F_{\rm rp}/F_{\rm self,g}]
\end{equation}
where $\tau_{\rm dyn}$ is the dynamical time of the galaxy component (rotation-supported or dispersion-supported), and $F_{\rm self,g}$ is the gravitational restoring force acting on gas in the galaxy midplane due to the galaxy's own self-gravity (evaluated at the half-mass radius of the galaxy, $r_{\rm hm}$). Removal of the galaxy \ISM\ will reduce the amount of new star formation occuring in satellites and so may affect $\lambda^*_{\rm Re/2}$;
 \item [Tidal stripping:] The \ISM\ and stars of the galaxy are stripped away via tidal forces. The tidal force is computed from the spherically-symmetric gravitational potential of the host halo. Mass loss rates of gas and stars from the galaxy are estimated using
\begin{equation}
 \dot{M}_{\rm (ISM|\star)} = - M_{\rm (ISM|\star)}/\tau_{\rm t},
\end{equation}
where
\begin{equation}
 \tau_{\rm t} = \tau_{\rm dyn} [\mathcal{F} r_{\rm hm} /F_{\rm self,t}]
\end{equation}
where $\mathcal{F}$ is the magnitude of the tidal field of the host halo, and $F_{\rm self,t}$ is the gravitational restoring force acting on the total (gas plus stellar) mass of the galaxy, evaluated at the galaxy half-mass radius. Removal of gas and stars (which may occur preferentially from the less tightly bound rotation-supported component) may affect $\lambda^*_{\rm Re/2}$;
 \item [Tidal stirring:] The tidal field of the host halo may drive bar instabilities in this disk of the satellite, which convert disk into spheroid and ``stir-up'' the spheroid by boosting its pseudo-angular momentum. Disk instabilities due to a disk's own self-gravity can be modelled in \glc\ using the stability parameter described by \cite{efstathiou_stability_1982}. Here, we modify that parameter\footnote{We retain the same functional form as used by \protect\cite{efstathiou_stability_1982} in the limit that the tidal field strength goes to zero, including the use of the maximum rotation speed as opposed to, for example, the rotation speed at the disk scale length. Given the flatness of galaxy rotation curves our results will be insensitive to the details of this choice. In particular, retaining the same choice as made by \protect\cite{efstathiou_stability_1982} avoids the need to recalibrate the critical value of the stability parameter.} to include a contribution from the tidal field of the host halo, which we assume can act to destabilize the disk:
\begin{equation}
 \epsilon = v_{\rm r, max} \left[{{\rm G} M_{\rm r} \over r_{\rm r}} + \hbox{max}(\mathcal{F} r_{\rm r}^2,0)\right]^{-1/2}.
\end{equation}
Here, $v_{\rm r,max}$ is the peak of the rotation curve, $M_{\rm r}$ is the total mass of the rotation-supported component, and $r_{\rm r}$ is the exponential scale length of the rotation-supported component. The rotation-supported component is deemed unstable to bar formation if $\epsilon < \epsilon_{\rm crit}$. For unstable systems, we assume that the bar causes material to transfer from the rotation to dispersion-supported components on a timescale of $\tau_{\rm dyn} (\epsilon_{\rm crit}-\epsilon_{\rm iso})/(\epsilon_{\rm crit}-\epsilon)$, where $\epsilon_{\rm iso}$ is the stability parameter of an isolated exponential disk (with no dark matter halo, and no tidal field). In this way the timescale equals the dynamical time of the rotation-supported component for a system at threshold and becomes shorter for more unstable systems. In addition, we assume that the tidal field torques the galaxy, increasing the pseudo-angular momentum of the dispersion-supported component at a rate $\dot{J} = \mathcal{F} M_{\rm s} r_{\rm s}^2$, where $r_{\rm s}$ is the half-mass radius of that component, which will cause the dispersion-supported component to expand and may affect its velocity dispersion.

The tidal response of a stellar system depends on the commensurability between the orbital time and the characteristic timescale of the perturbation. The latter can be simply measured by the time spent near pericenter, $t_{\rm peri}$, where the tidal shock occurs. The reponse of the system is strongest when the ratio between 
$t_{\rm peri}$ and the internal orbital time $t_{\rm orb}$ is of order unity or smaller. In this case, which is well described by the impulse approximation, the internal kinetic energy of the system can increase substantially, which results in a higher velocity dispersion and eventually mass stripping. On the contrary, when $t_{\rm peri}/t_{\rm orb} > 1$, the internal kinetic energy of the system is less affected by the tidal perturbation, and the object 
remains closer to its original equilibrium configuration \citep{gnedin_tidal_1999}. In this case adiabatic corrections are required to describe the tidal response of the system \citep{gnedin_effects_1999}. While most studies have focused on exploring the two regimes for tidal mass loss of halos or galaxies \citep[e.g.]{kazantzidis_density_2004} rather than for tidally induced non-axisymmetric instabilities, numerical results point to weaker bars and a reduced effect of the transformation by tidal stirring whenever the response is more adiabatic due to a steeper halo or stellar density profile or owing to a particular orbital configuration \citep{mayer_metamorphosis_2001,kazantzidis_tidal_2013}. We have checked that nearly 70\% of the galaxies in our sample actually evolve in the impulsive rather than adiabatic regime over the entire period for which we assume the bar instability is taking place. Therefore, while our model does not include adiabatic corrections, this suggests that our results on the efficiency of the transformation, despite the simplicity of our model, should be robust.
\end{description}

\begin{table}
 \caption{Values of $(v/\sigma)^*$ for four model galaxies evolved in the cluster potential of \protect\cite{mastropietro_morphological_2005} for 5~Gyr. Results from our semi-analytic calculation are shown in the second column, while results from the N-body calculations of \protect\cite{mastropietro_morphological_2005} are shown in the third column.}
 \label{tb:calibration}
 \begin{center}
 \begin{tabular}{lll}
 \hline
             & \multicolumn{2}{c}{\boldmath{$v/\sigma$} {\bf at 5~Gyr}} \\
 {\bf Model} & {\bf SAM} & {\bf N-body} \\
 \hline
 GAL2 & 0.013 & 0.075 \\
 GAL5 & 1.05  & 1.10 \\
 GAL6 & 0.04  & 0.10 \\
 GAL7 & 1.53  & 1.35 \\
 \hline
 \end{tabular}
 \end{center}
\end{table}

For the tidal stirring model, we have one free parameter, namely the stability threshold, $\epsilon_{\rm crit}$. \cite{efstathiou_stability_1982} find that a value of $1.1$ provides the best match to their simulations. Here, we calibrate the value of $\epsilon_{\rm crit}$ by comparing to the simulations of \cite{mastropietro_morphological_2005}. We set up analogs of their simulations in \glc\ and evolve four satellites (corresponding to their models GAL2, GAL5, GAL6, and GAL7) for 5~Gyr. \cite{mastropietro_morphological_2005} do not measure the $\lambda^*_{\rm Re/2}$ statistic, but instead measure $(v/\sigma)$, using the peak of the rotation curve for $v$ and the line-of-sight dispersion averaged within the half-mass radius of the galaxy for $\sigma$. We estimate this quantity from our model using a mass-weighted average of the rotation velocity of the rotation-supported component (measured at the H-band half-light radius of the galaxy) and the velocity dispersion of the dispersion-supported (specifically the luminosity-weighted line-of-sight velocity dispersion averaged within the H-band half-light radius of the galaxy) as follows:
\begin{equation}
{v\over\sigma} = \left({v^2 M_{\rm d} \over \sigma^2 M_{\rm s}}\right)^{1/2},
\end{equation}
where $M_{\rm d}$ and $M_{\rm s}$ are the masses of the disk and spheroid components respectively.

Comparing to the results in Fig.~11 of \cite{mastropietro_morphological_2005}, we find good agreement (see Table~\ref{tb:calibration}) for $\epsilon_{\rm crit} = 1.12$ (which is very close to the preferred value found by \citealt{efstathiou_stability_1982}) and so adopt this value for all calculations. This also illustrates that our simple model correctly captures the dependence of $v/\sigma$ on pericentric distance.

\begin{figure}[t]
 \includegraphics[width=85mm]{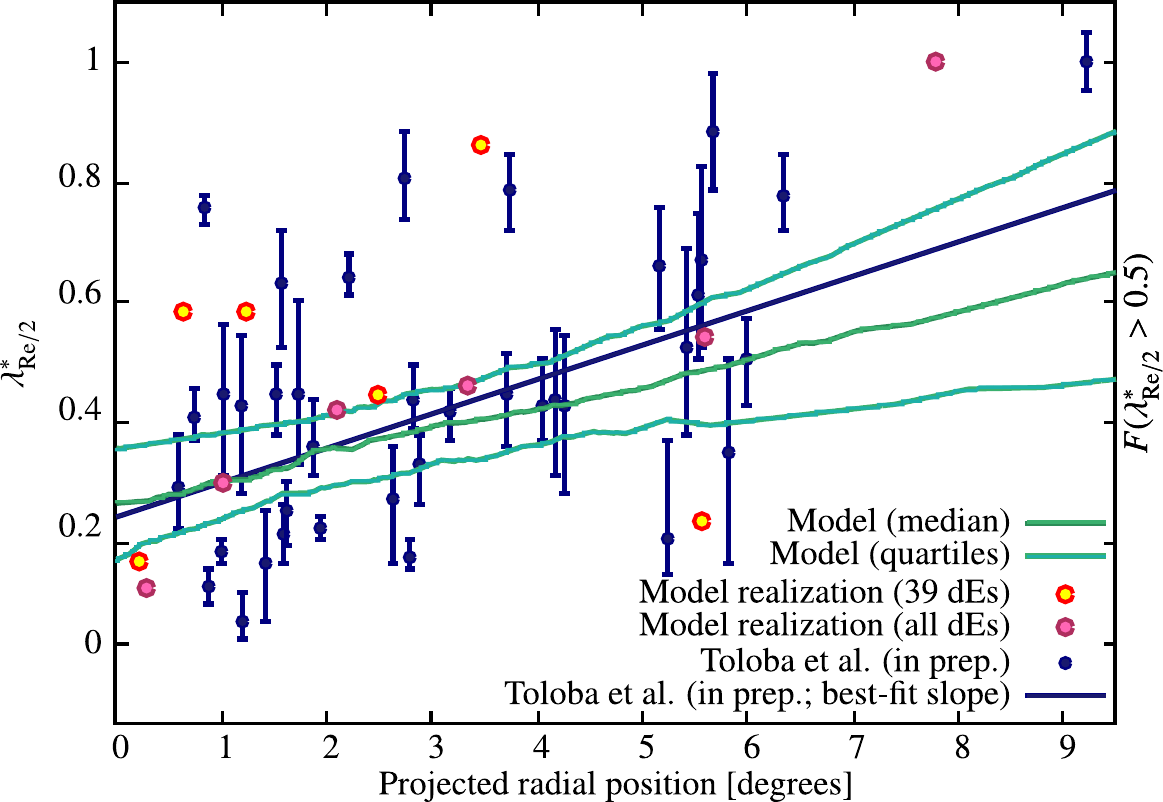}
 \caption{The dynamical state, $\lambda^*_{\rm Re/2}$, of dwarf early-type galaxies from a single Virgo cluster realization in our tidal stirring model is shown as a function of the projected radial position of the galaxy from the cluster center. Yellow points show the fraction of model galaxies with $\lambda^*_{\rm Re/2} > 0.5$, $F(\lambda^*_{\rm Re/2} > 0.5)$, in adaptive bins chosen to contain 8 galaxies each from a randomly selected sample of 39 dwarf early-types from a randomly selected cluster realization. Purple points show the $F(\lambda^*_{\rm Re/2} > 0.5)$ averaged over six adaptive bins (chosen to contain equal numbers of galaxies) when all dwarf early-types in the realization are used. For comparison, blue points show the observations of \cite{toloba_stellar_2014a}, with the blue line indicating the best-fit slope to these data. The dark green line shows the median model relation in the mean value of $\lambda^*_{\rm Re/2}$ vs. radius (over all 100 cluster realizations) while the light green lines indicate the $25^{\rm th}$ and $75^{\rm th}$ percentiles of the model distribution.}
 \label{fig:realization}
\end{figure}

From each realization we randomly select 39 dwarf early-type galaxies (the same number as in the observed sample of \citealt{toloba_stellar_2014a}) from the set of galaxies having $-18 < M_{\rm B} < -14$ and at least 25\% of their stellar mass in their dispersion-supported component. Each satellite's orbit is projected onto the plane of the sky (assuming an isotropic distribution of angular momentum vectors and random orbital phase) to determine the observed angular distance of satellite from cluster center\footnote{We assume a distance to Virgo of $16.5$~Mpc \protect\citep{mei_acs_2007}.}. We note that individual model points cluster along $\lambda^*_{\rm Re/2}=0$ and 1, indicating that model galaxies tend to be either fully rotationally supported or fully dispersion supported, while observed galaxies are distributed more uniformly over $\lambda^*_{\rm Re/2}$. This difference likely arises from two effects. First, the model galaxies (and our model for tidal stirring) are clearly approximations that do not capture the full dynamics of the real systems. Second, the model galaxy $\lambda^*_{\rm Re/2}$ values are error free---adding errors would tend to scatter the points away from $\lambda^*_{\rm Re/2}$. Given the simplicity of the current modelling we do not believe a detailed attempt to make mock observations of the models (and thereby assign errors) is warranted\footnote{As we have noted, if we were to explicitly take into account adiabatic corrections the tidal response might be weaker, which would likely add to the scatter in the model points. Additionally, tidally stirred dwarfs will retain some degree of triaxiality for several Gyr, having residual rotation in more than one axis  \protect\citep{mastropietro_morphological_2005} which would also contribute to scatter in the model points.}.

An example of the resulting distribution of $\lambda^*_{\rm Re/2}$ vs. angular radius for the tidal stirring model is shown in Figure~\ref{fig:realization}. Yellow points show the fraction\footnote{Given that model galaxies cluster around $\lambda^*_{\rm Re/2}=0$ and 1, plotting this fraction is more instructive than plotting the mean of $\lambda^*_{\rm Re/2}$.} of model galaxies with $\lambda^*_{\rm Re/2} > 0.5$, $F(\lambda^*_{\rm Re/2} > 0.5)$, in adaptive bins chosen to contain 8 galaxies each from a randomly selected sample of 39 dwarf early-types. Purple points show $F(\lambda^*_{\rm Re/2} > 0.5)$ in six adaptive bins (chosen to contain equal numbers of galaxies) when all dwarf early-types in the realization are used. For comparison, blue points show the observational data of \cite{toloba_stellar_2014a}, with the blue line indicating the best-fit slope to these data.

To quantify the radial trend in $\lambda^*_{\rm Re/2}$ we find the slope, $\alpha$, of the best-fit linear relation between {\boldmath $F(\lambda^*_{\rm Re/2} > 0.5)$} and projected radius. Given the simplicity of the model, this provides a simple, yet robust measure of the variation in $\lambda^*_{\rm Re/2}$ with radius in the cluster which can be reliably compared to the observational measurement. In Figure~\ref{fig:realization} the dark green line shows the median of the fitted linear relations over all 100 cluster realizations, while the the light green lines indicate the $25^{\rm th}$ and $75^{\rm th}$ percentiles of the model distribution.

\section{Results}\label{sec:Results}

\begin{figure}[t]
 \includegraphics[width=85mm]{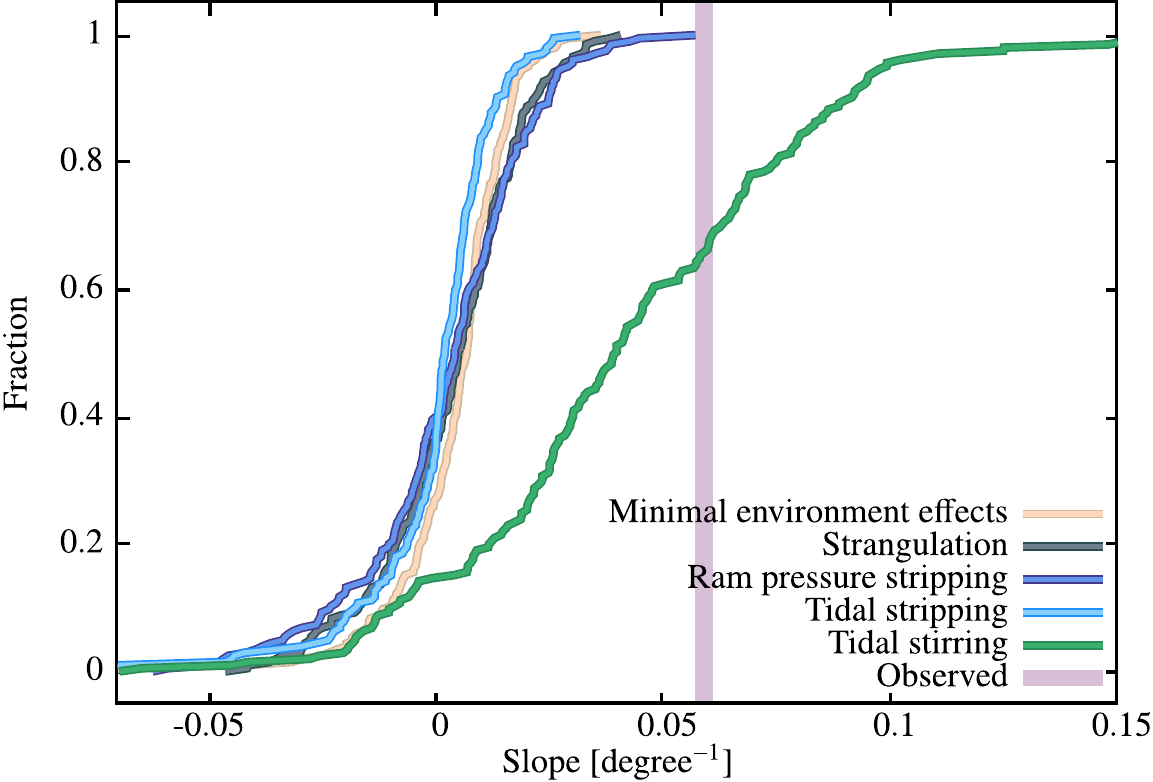}
 \caption{The cumulative distribution of slopes, $\alpha$, of the best-fit linear relation to the mean $\lambda^*_{\rm Re/2}$ as a function of projected radius. Colored lines show the results for our different models. The vertical band indicates the value and $\pm 1\sigma$ errors measured from the observational data of \protect\cite{toloba_stellar_2014a}.}
 \label{fig:slopes}
\end{figure}

Figure~\ref{fig:slopes} shows the cumulative distribution of slopes, $\alpha$, in the best-fit linear relation between $\lambda^*_{\rm Re/2}$ and projected radius, measured from 100 Virgo cluster realizations for each model. As these slopes are measured using the same number of galaxies as were used in the observational sample, random fluctuations in $\alpha$ due to small number statistics should be matched between model and observations. Line colors correspond to different models as shown in the figure. The vertical band indicates the value of $\alpha$ measured from \protect\cite{toloba_stellar_2014a}. We expect that, for a plausible model, the observed slope should correspond to a probability neither too small, nor too high. We find that none of our 100 realizations of the ``minimal environmental effects'' model have $\alpha$ as large as that observed. This same result is found for strangulation, ram pressure and tidal stripping models. The tidal stirring model shows a markedly different result, with $P(>\alpha_{\rm obs})= 34\%$.

\section{Conclusions}\label{sec:Conclusions}

We have performed calculations of the effects of several different environmental effects on the dynamics of dwarf early-type galaxies in the Virgo cluster to assess the strength of radial trends in $\lambda^*_{\rm Re/2}$. Quantifying the statistics of these trends requires realizations of large numbers of dEs in the cluster and, furthermore, many realizations of the entire cluster. To achieve this, we employ a semi-analytic model of galaxy formation which permits rapid construction of many cluster realizations. While this technique cannot capture the full details of dynamical evolution as would an N-body or hydrodyamic simulation, we have demonstrated that its results are in good agreement with extant simulations.

We find that a model in which there are no environmental effects on cluster galaxies (apart from their being cut off from any gas accretion from the \IGM) never produces a slope in the mean $\lambda^*_{\rm Re/2}$ with projected radius which exceeds that measured by \protect\cite{toloba_stellar_2014a} in any of our 100 cluster realizations. Models including ram pressure or tidal stripping, or strangulation (rapid removal of the hot gas atmosphere associated with dEs once they become cluster members) similarly fail to produce any realizations with slopes consistent with that which is observed. However, a model in which tidal stirring drives the evolution of $\lambda^*_{\rm Re/2}$ is able to produce slopes exceeding that observed 34\% of the time.

These results strongly rule out the null hypothesis that there are no environmental effects at work in shaping the observed trend of $\lambda^*_{\rm Re/2}$ with radius. Given the present observations, the only viable model is one which includes the effects of tidal stirring. This conclusion could be tested through similar observations of other clusters---our tidal stirring model would predict that approximately 85\% of Virgo-like clusters should show radial trends in which $\lambda^*_{\rm Re/2}$ increases with radius.

\protect\cite{toloba_stellar_2014a} chose to measure $\lambda^*$ within Re$/2$ in order to minimize errors in their measurement. We note that, had the observational measurement been made within Re rather than Re$/2$ the measured slope of $\lambda^*$ with radius would change from $\alpha=0.06\pm0.01$ to $\alpha=0.07\pm0.1$ \protect\citep{toloba_stellar_2014a} which would not affect our conclusions.

Our modelling of environmental effects has been kept purposely simplistic in this current work. Future work could incorporate more detailed modelling of satellite orbits \citep{taylor_dynamics_2001,benson_effects_2002,taylor_evolution_2004} and perform more extensive calibrations of our semi-analytic algorithms for environmental effects against N-body/hydrodyamic simulations. Coupled with expanded observational datasets, this approach has the potential to place strong constraints on the nature of environmental effects acting on cluster galaxies.

\acknowledgments

E.T. acknowledges the financial support of the Fulbright program jointly with the Spanish Ministry of Education. P.G. acknowledges support from NSF grant AST-1010039.

\bibliographystyle{apj}
\bibliography{dwarfEllipticalsAccented}

\end{document}